\documentclass[
 reprint,
superscriptaddress,
showpacs,showkeys,preprintnumbers,
 amsmath,amssymb,
aps,
pra,
]{revtex4-1}

\usepackage{graphicx}
\usepackage{dcolumn}
\usepackage{bm}
\usepackage{hyperref}

\usepackage{dsfont}

\begin{document}


\title{Quaternionic Quantum Dynamics on Complex Hilbert Spaces}

\author{Matthew A.\ Graydon}
\email{mgraydon[at]perimeterinstitute.ca}
\affiliation{Perimeter Institute for Theoretical Physics, 31 Caroline St.\ N., Waterloo, Ontario N2L 2Y5, Canada}
\affiliation{Department of Physics $\&$ Astronomy, University of Waterloo, Waterloo, Ontario N2L 3G1, Canada}
\date{\today}
\begin{abstract}
We consider a quaternionic quantum formalism for the description of quantum states and quantum dynamics. We prove that generalized quantum measurements on physical systems in quaternionic quantum theory can be simulated by usual quantum measurements with positive operator valued measures on complex Hilbert spaces. Furthermore, we prove that quaternionic quantum channels can be simulated by completely positive trace preserving maps on complex matrices. These novel results map all quaternionic quantum processes to algorithms in usual quantum information theory.  
\end{abstract}

\pacs{03.65.-w, 03.67.Ac}
\maketitle


\section{\label{sec:level1}Introduction}

Physical theories prescribe probability calculi for computing measurement outcome expectations and state assignments for physical systems.  
The probability calculi prescribed by classical theories of physics are fundamentally different from the probability calculus prescribed by quantum mechanics. For instance, Birkhoff and von Neumann pointed out that classical experimental propositions regarding physical systems form Boolean algebras; whereas, quantum experimental propositions comprise nondistributive orthomodular lattices \cite*{vonNeumann36}. Moreover, Feynman emphasized that the classical Markovian law of probability composition fails to hold in the description of general quantum mechanical phenomena \cite*{Feynman48}\cite*{Adler95}. Instead, quantum  \textit{probability amplitudes} superimpose. These essential features are not unique to the calculus prescribed by usual quantum mechanics as a theory over the complex field -- they are enjoyed in quantum theories formulated over any of the associative normed division rings $\mathbb{R}$, $\mathbb{C}$, or $\mathbb{H}$.\\
\indent What, then, \textit{does} distinguish quantum theories formulated over $\mathbb{R}$ or $\mathbb{H}$ from usual complex quantum mechanics (\textsc{cqm})? In the case of real quantum theory (\textsc{rqt}) \cite*{Stueckelberg60}, multipartite systems are endowed with some rather unusual properties. For example, in \textsc{rqt}, there exist states associated with $n$-partite systems for which every subsystem is maximally entangled with each of the other subsystems, where $n$ can be arbitrarily large \cite*{Wootters10}. Furthermore, \textsc{rqt} is not a locally tomographic theory -- it is instead a bilocally tomographic theory \cite*{Hardy10}. These observations point to aspects of \textsc{rqt} that cannot be realized within the usual \textsc{cqm} framework. However, the evolution and measurement of a multipartite complex quantum state under discrete or continuous evolution in \textsc{cqm} \textit{can} be simulated using states and operators in \textsc{rqt} \cite*{McKague09a}.\\
\indent In the case of quaternionic quantum theory (\textsc{qqt}) \cite*{Finkelstein59}\cite*{Finkelstein62}\cite*{Finkelstein63}, the very notion of `independent subsystems' is ill defined. In fact, quaternion-linear tensor products of quaternionic modules \textit{do not exist} \cite*{Razon91}. This constitutes a significant obstacle for the development of a consistent definition of local quaternionic operations, and it has been argued that one is actually prevented from speaking of absolutely independent systems in \textsc{qqt} \cite*{Finkelstein62}. These peculiar features may set \textsc{qqt} apart from usual \textsc{cqm}. Nevertheless, in the context of $1$-dimensional quantum wave mechanics, it has been shown that \textsc{cqm} is consistent with \textsc{qqt}. Specifically, one can recreate the entire structure of $1$-dimensional complex quantum wave mechanics inside $1$-dimensional quaternionic quantum wave mechanics \cite*{Nash92}. Conversely, the experimental propositions in \textsc{qqt} that commute with a fixed anti-Hermitian unitary operator are isomorphic to the experimental propositions of \textsc{cqm} \cite*{Finkelstein62}. In the context of quantum information processing involving unitary transformations and projective measurements, it has been shown that circuits acting on $n$ $2$-dimensional quaternionic systems can be simulated by circuits acting on $n+1$ qubits \cite*{Fernandez03}.\\
\indent In this paper, we consider a \textit{generalized} formulation of dynamics in \textsc{qqt}, rather than only considering the restricted class of quantum processes treated in \cite*{Fernandez03}. We treat generalized quaternionic quantum measurements as positive operator valued measures on quaternionic modules, and we treat quaternionic quantum channels as completely positive trace preserving quaternionic maps. Given an arbitrary $d$-dimensional quaternionic quantum state $\rho$ for a physical system $\mathfrak{S}$, we show that a generalized quaternionic quantum measurement $\mathfrak{M}_{\mathbb{H}}$ on $\mathfrak{S}$ can be simulated by a complex quantum measurement $\mathfrak{M_{\mathbb{C}}}$ on $\mathfrak{S}$ with an associated $2d$-dimensional complex quantum state $\sigma$. We also show that any quaternionic quantum channel can be simulated by a completely positive trace preserving map in \textsc{cqm}.\\ 
\indent The remainder of this paper is  structured as follows. In section II, we review, for the reader's convenience, prerequisite material concerning quaternions and quaternionic linear algebra.  In section III, we introduce a quaternionic quantum formalism, and we prove a Gleason-type theorem dictating the quaternionic Born rule for calculating probabilities for outcomes of generalized measurements in \textsc{qqt}. In section IV, we exhibit quaternionic quantum dynamics as complex quantum dynamics on complex Hilbert spaces. Finally, we conclude in Section V. 

\section{\label{sec:level2}Quaternionic Algebra}
\subsection{\label{sec:level3}Quaternions}
The quaternions were first discovered by Hamilton \cite*{Hamilton44}. We express $h\in\mathbb{H}$ as $h=1h_{0}+ih_{1}+jh_{2}+kh_{3}$ in terms of its constituents $h_{r}\in\mathbb{R}\;\forall r\in\{0,1,2,3\}$, and the quaternion basis elements $\{1,i,j,k\}$, which obey  
\begin{equation}
i^{2}=j^{2}=k^{2}=ijk=-1 \text{.}
\label{basisElements}
\end{equation}
$\mathbb{H}$ is an abelian group with respect to addition defined via $h+h^{'}=1(h_{0}+h^{'}_{0})+i(h_{1}+h^{'}_{1})+j(h_{2}+h^{'}_{2})+k(h_{3}+h^{'}_{3})$, and a monoid with respect to noncommutative multiplication defined via 
\begin{eqnarray}
hh^{'} = & 1(h_{0}h^{'}_{0}-h_{1}h^{'}_{1}-h_{2}h^{'}_{2}-h_{3}h^{'}_{3})+ \nonumber \\ 
 & i(h_{0}h^{'}_{1}+h_{1}h^{'}_{0}+h_{2}h^{'}_{3}-h_{3}h^{'}_{2})+ \nonumber \\ 
 & j(h_{0}h^{'}_{2}+h_{2}h^{'}_{0}-h_{1}h^{'}_{3}+h_{3}h^{'}_{1})+ \nonumber \\ 
 & k(h_{0}h^{'}_{3}+h_{3}h^{'}_{0}+h_{1}h^{'}_{2}-h_{2}h^{'}_{1})\text{,}\;\;\;
\end{eqnarray}
$\forall h,h^{'}\in\mathbb{H}$. Quaternion addition and multiplication are distributive in the sense that  $h(h^{'}+h^{''})=hh^{'}+hh^{''}$, and $(h+h^{'})h^{''}=hh^{''}+h^{'}h^{''}$ $\forall h,h^{'},h^{''}\in\mathbb{H}$. The quaternionic conjugation operation $h \rightarrow \overline{h}$ taking $\{1,i,j,k\}\rightarrow\{1,-i,-j,-k\}$ is an involutory anti-automorphism inducing a multiplicative norm $|h|=(h\overline{h})^{\frac{1}{2}}$ on $\mathbb{H}$. 

The compact symplectic group Sp(1) of unit-norm quaternions is isomorphic to SU(2), which can be seen from viewing $\varphi\in$ Sp(1) as $\varphi=\gamma_{1}+\gamma_{2}j$ in terms of \begin{eqnarray}
\gamma_{1}=1\varphi_{0}+i\varphi_{1}\in\mathbb{C} \label{symCorRep1} \text{,}\\
\gamma_{2}=1\varphi_{2}+i\varphi_{3}\in\mathbb{C} \text{,}
\label{symCorRep2}
\end{eqnarray}
so that $\varphi\overline{\varphi}=1\implies |\gamma_{1}|^{2}+|\gamma_{2}|^{2}=1$. Next, by defining $f:$ Sp(1) $\rightarrow$ SU(2) such that
\begin{equation}
f(\varphi)=\begin{bmatrix}\;\;\;\gamma_{1} &\gamma_{2} \vspace{0.5ex}\\ -\overline{\gamma_{2}} & \overline{\gamma_{1}}\end{bmatrix}\text{,}
\label{symRep}
\end{equation}
it is clear that $f$ is a bijection and that $f(\varphi_{1}\varphi_{2})=f(\varphi_{1})f(\varphi_{2})$, establishing that Sp(1) $\cong$ SU(2). More generally, one has that Sp(d) $\cong$ U(2d, $\mathbb{C}$) $\cap$ Sp(2d, $\mathbb{C}$), where Sp(d) $\cong$ U(d, $\mathbb{H}$) is the group of $d\times d$ unitary quaternionic matrices \cite*{Fulton91}. There are, however, subtle distinctions between quaternionic and complex matrix algebras due to the noncommutativity of quaternion multiplication. 

\subsection{\label{sec:level4}Quaternionic Modules and Matrices}

For an excellent review of quaternionic linear algebra, we refer the reader to \cite*{Zhang97}. In this paper, we adopt the convention wherein the Cartesian product $\mathbb{H}^{d}$ is taken as a \textit{right} quaternionic module. We equip $\mathbb{H}^{d}$ with the standard symplectic inner product $\langle \cdot|\cdot\rangle:\mathbb{H}^{d}\rightarrow\mathbb{H}$ defined via $\langle \phi|\chi\rangle=\sum_{r=1}^{d}\overline{\phi_{r}}\chi_{r}$ $\forall \phi, \chi\in\mathbb{H}^{d}$, where $\phi_{r}$ and $\chi_{r}$ denote the projections of $\phi$ and $\chi$ onto elements of a basis for $\mathbb{H}^{d}$. Furthermore, we adopt the convention wherein linear operators on $\mathbb{H}^{d}$ act as elements of $M_{p,d}(\mathbb{H})$ -- the set of $p\times d$ quaternionic matrices -- from the \textit{left}, as usual. Stated explicitly, if $A\in M_{p,d}(\mathbb{H})$ with entries $[A]_{rs}$ and $\phi\in\mathbb{H}^{d}$, then our conventions imply that $A\phi$ is computed as
\begin{equation}
A\phi=\sum_{r=1}^{p}\sum_{s=1}^{d}\sum_{t=1}^{d}|r\rangle A_{rs} \langle s|t\rangle\phi_{t}=\sum_{r=1}^{p}\sum_{s=1}^{d}|r\rangle A_{rs}\phi_{s}\text{,} 
\label{matrixMult}
\end{equation}
where the last equality in \eqref{matrixMult} follows in general if and only if $s$ and $t$ are elements of an orthonormal basis. 

Following Finkelstein et al.\ \cite*{Finkelstein59}, we define the \textit{trace} of $A\in M_{d,d}(\mathbb{H})$ with respect to a basis $\Omega=\left\{\omega_{1},\omega_{2},\dots,\omega_{d}\right\}$ for $\mathbb{H}^{d}$ as $\mathrm{tr}(A)=\mathrm{Re}\left(\sum_{r=1}^{d}\langle\omega_{r}|A\omega_{r}\rangle\right)$. It follows that the trace of $A$ is independent of $\Omega$. It also follows that the cyclic property of the trace holds: $\mathrm{tr}(ABC)=\mathrm{tr}(CAB)$ $\forall A,B,C\in M_{d,d}(\mathbb{H})$. $M_{p,d}(\mathbb{H})$ admits an involution $^{*}$ defined such that $[A^{*}]_{rs}=\overline{[A]_{sr}}$. When $p=d$, $\langle\phi|A\chi\rangle=\langle A^{*}\phi|\chi\rangle$, and we denote the set of self-adjoint quaternionic matrices satisfying $A=A^{*}$ by $M_{d,d}(\mathbb{H})_{sa}$. Given our conventions, the spectral theorem holds for self-adjoint quaternionic matrices \cite*{Adler95}. As usual, we say that $A\in M_{d,d}(\mathbb{H})$ is \textit{unitary} when $AA^{*}=\mathds{1}_{\mathbb{H}^{d}}$, and we say that $A\in M_{d,d}(\mathbb{H})$ is \textit{positive semi-definite} when $\langle\phi|A\phi\rangle\ge0$ $\forall\phi\in\mathbb{H}^{d}$. We state without proof that positive semi-definiteness implies self-adjointness for elements of $M_{d,d}(\mathbb{H})$. We equip $M_{d,d}(\mathbb{H})_{sa}$ with the symmetric positive-definite $\mathbb{R}-$bilinear form $(\cdot,\cdot): M_{d,d}(\mathbb{H})_{sa}\times M_{d,d}(\mathbb{H})_{sa} \rightarrow \mathbb{R}$ defined via $(A,B)=\mathrm{tr}(AB)$. For the remainder of this paper we shall view $M_{d,d}(\mathbb{H})_{sa}$ as a real vector space. On that view, $\mathrm{tr}(AB)$ is an \textit{inner product} on the real vector space $M_{d,d}(\mathbb{H})_{sa}$ inducing the norm $|A|=\sqrt{\mathrm{tr}(A^{2})}$.

\subsection{\label{sec:level5}\texorpdfstring{Embedding $M_{p,d}(\mathbb{H})$ into $M_{2p,2d}(\mathbb{C})$}{Embedding Mp,d(H) into M2p,2d(C)}}
Let $A\in M_{p,d}(\mathbb{H})$  be a $p\times d$ quaternionic matrix with $A=\Gamma_{1}+\Gamma_{2}j$, where $\Gamma_{1},\Gamma_{2}\in M_{p,d}(\mathbb{C})$ are obtained by decomposing the matrix elements $[A]_{rs}$ according to \eqref{symCorRep1} and \eqref{symCorRep2}. In analogy with \eqref{symRep}, we define the embedding $\psi_{p,d}:M_{p,d}(\mathbb{H})\rightarrow M_{2p,2d}(\mathbb{C})$ via
\begin{equation}
\psi_{p,d}\left(A\right)=
\begin{bmatrix}
  \;\;\;\Gamma_{1} & \Gamma_{2}\; \vspace{0.5ex}\\
  -\overline{\Gamma_{2}} & \overline{\Gamma_{1}}\;
\end{bmatrix}
\text{.}
\label{embed}
\end{equation}
If $a,a'\in\mathbb{R}$, $A,A'\in M_{p,d}(\mathbb{H})$, and $B\in M_{d,q}(\mathbb{H})$, then it follows from \eqref{embed} that
\begin{equation}
\psi_{p,d}(aA'+a'A')=a\psi_{p,d}\left(A\right)+a'\psi_{p,d}\left(A'\right)\text{,}
\label{rLin}
\end{equation}
\begin{equation}
\psi_{p,d}\left(A\right)\psi_{d,q}\left(B\right)=\psi_{p,q}\left(AB\right) \text{,}
\label{homo}
\end{equation}
\begin{equation}
\psi_{d,p}\left(A^{*}\right)=\psi_{p,d}\left(A\right)^{*} \text{.}
\label{star}
\end{equation}
It is also readily verified that $\psi_{p,d}$ is an injection. Furthermore, when $p=d=q$, $\psi_{d,d}$ is the usual injective $^{*}$-homomorphism from $M_{d,d}(\mathbb{H})$ into $M_{2d,2d}(\mathbb{C})$ pointed out by Farenick and Pidkowich in \cite*{Farenick03}.

\section{\label{sec:level6}Quaternionic Quantum Formalism}

\subsection{\label{sec:level7}States, Evolution, and Measurement}
In \textsc{qqt}, a quantum state for a $d$-dimensional physical system is associated with a unit-trace positive semi-definite matrix $\rho\in M_{d,d}(\mathbb{H})_{sa}$. We will assume that the time-evolution of a quantum state $\rho$ is governed by a quantum channel $\Phi$ whose action is defined by a completely positive trace preserving quaternionic map \cite*{Kossakowski00}. On that assumption, we take $\Phi(\rho)=\sum_{r=1}^{n}A_{r}\rho A_{r}^{*}$, where $A_{r}\in M_{p,d}(\mathbb{H})$ are such that $\sum_{r=1}^{n}A_{r}A_{r}^{*}=\mathds{1}_{\mathbb{H}^{p}}$, and where $n\in \mathbb{Z}_{+}$. We will associate a quaternionic quantum measurement device $\mathfrak{M}_{\mathbb{H}}$ with a positive operator valued measure on $\mathbb{H}^{d}$ whose values are $\{E_{1},\dots,E_{m}\}$, $m\in\mathbb{Z}_{+}$, such that $E_{r}\in M_{d,d}(\mathbb{H})$ are positive semi-definite and $\sum_{r=1}^{m}E_{r}=\mathds{1}_{\mathbb{H}^{d}}$. Each $E_{r}$ corresponds to a measurement outcome that may occur with a probability given by the Born rule.

\subsection{\label{sec:level8}The Born Rule in QQT}
For dimension $d\ge3$,  the Born rule for calculating probabilities for outcomes of projection valued measurements in usual \textsc{cqm} was derived by Gleason \cite*{Gleason57}. Gleason's result carries over to \textsc{rqt} and \textsc{qqt} \cite*{Buhagiar09}. Caves et al.\ extended Gleason's result in a noncontextual setting to cover quantum measurements associated with positive operator valued measures on complex Hilbert spaces for all dimensions $d\ge2$ \cite*{Caves04}. 

We will now proceed to show that the result given by Caves et al.\ carries over to \textsc{qqt}. Let us denote by $\mathcal{E}(\mathbb{H}^{d})$ the set of all \textit{quaternionic quantum effects} -- that is, the set of all positive semi-definite linear operators $E$ on $\mathbb{H}^{d}$ admitting $\mathrm{tr}(E^{2})\le d$. We define a \textit{quaternionic frame function} as any map $f:\mathcal{E}(\mathbb{H}^{d})\rightarrow[0,1]$ satisfying
\begin{equation}
\sum_{E_{r}\in X}f(E_{r})=1\text{,}\;\forall X=\Big\{E_{r}\in\mathcal{E}(\mathbb{H}^{d})\;\Big|\;\sum_{r}E_{r}=\mathds{1}_{\mathbb{H}^{d}}\Big\}.
\end{equation}
For every frame function $f$, there exists a unique unit-trace positive semi-definite $\rho\in M_{d,d}(\mathbb{H})_{sa}$ such that 
\begin{equation}
f(E)=(E,\rho)=\mathrm{tr}(E\rho).
\label{bornRule} 
\end{equation}
This is the Born rule for calculating probabilities for outcomes of generalized measurements in \textsc{qqt}. For the proof, note that quaternionic quantum effects admit a spectral resolution in terms of real eigenvalues and orthogonal eigenprojectors. As a result, the proof given by Caves et al.\ in the complex case can almost literally be transfered to the quaternionic case, and we encourage the reader to consult \cite*{Caves04} for details. In particular, one can establish $\mathbb{R}-$linearity of $f$ on $M_{d,d}(\mathbb{H})_{sa}$. Now, let $\{\Upsilon_{1},\dots,\Upsilon_{d(2d-1)}\}$ be an orthonormal basis for $M_{d,d}(\mathbb{H})_{sa}$. We can expand any effect $E$ as a linear combination of the $\Upsilon_{r}$ in terms of coefficients $(\Upsilon_{r},E)$. Also, there  exists a unique operator $\rho$ that we can expand as a linear combination of the $\Upsilon_{r}$ in terms of coefficients $f(\Upsilon_{r})$. It follows that $f(E)=(E,\rho)$. The operator $\rho$ is positive semi-definite, which is verified by letting $E=|\phi\rangle\langle\phi|$ for arbitrary $\phi\in\mathbb{H}^{d}$. We also have that $\rho$ is unit-trace, which follows from the observation that $\mathrm{tr}(\rho)=(\rho,\mathds{1}_{\mathbb{H}^{d}})=\left(\rho,\sum_{E_{r}\in X}E_{r}\right)=\sum_{E_{r}\in X}f(E_{r})=1$, finishing the proof. It is worth mentioning that these arguments would fail to hold if we had used the standard Hilbert-Schmidt inner product on $M_{d,d}(\mathbb{H})_{sa}$,  which is not  real-valued in general.

\section{\label{sec:level9}Complex Simulations of Quaternionic Quantum Dynamics}

\subsection{\label{sec:level0}Inner Product Correspondence}
Before we show that quaternionic quantum dynamics can be simulated by complex quantum dynamics, it will be useful to establish the following correspondence between our inner product on $M_{d,d}(\mathbb{H})_{sa}$ and the usual Hilbert-Schmidt inner product on $M_{2d,2d}(\mathbb{C})_{sa}$:
\begin{equation}
\mathrm{tr}(AB) = \textstyle{\frac{1}{2}}\mathrm{tr}\Big(\psi_{d,d}(A)\psi_{d,d}(B)\Big)\;\;\forall A,B \in M_{d,d}(\mathbb{H})_{sa}\text{.}
\label{claim}
\end{equation}
For the proof, we expand $A=\Gamma_{1}+\Gamma_{2}j$ and $B=\Lambda_{1}+\Lambda_{2}j$ in terms of complex self-adjoint $\Gamma_{1}=\Gamma_{1}^{*}$ and $\Lambda_{1}=\Lambda_{1}^{*}$, and complex antisymmetric $\Gamma_{2}=-\Gamma_{2}^{\mathrm{T}}$ and $\Lambda_{2}=-\Lambda_{2}^{\mathrm{T}}$. Expanding the LHS of \eqref{claim} we get

\begin{eqnarray}
 & \underbrace{\textstyle{\frac{1}{2}}\mathrm{tr}\big(\Gamma_{1}\Lambda_{1}+\Lambda_{1}\Gamma_{1}\big)}_{\mathtt{\alpha}}+\underbrace{\textstyle{\frac{1}{2}}\mathrm{tr}\big(\Gamma_{2}j\Lambda_{2}j+\Lambda_{2}j\Gamma_{2}j\big)}_{\mathtt{\beta}} \nonumber \\
 & + \underbrace{\textstyle{\frac{1}{2}}\mathrm{tr}\big(\Gamma_{1}\Lambda_{2}j+\Lambda_{1}\Gamma_{2}j+\Gamma_{2}j\Lambda_{1}+\Lambda_{2}j\Gamma_{1}\big)}_{\mathtt{\delta}}\text{,}
\label{LHS}
\end{eqnarray}
whereas expanding the RHS of \eqref{claim} we get
\begin{equation}
\underbrace{\textstyle{\frac{1}{2}}\mathrm{tr}\left(\Gamma_{1}\Lambda_{1}+\overline{\Gamma_{1}}\;\overline{\Lambda_{1}}\right)}_{\mathtt{\alpha}'}+\underbrace{\textstyle{\frac{1}{2}}\mathrm{tr}\left(-\Gamma_{2}\overline{\Lambda_{2}}-\overline{\Gamma_{2}}\Lambda_{2}\right)}_{\mathtt{\beta}'}\text{.}
\label{RHS}
\end{equation}
It is not hard to see that $\mathtt{\alpha}=\mathtt{\alpha}'$, $\mathtt{\beta}=\mathtt{\beta}'$, and $\mathtt{\delta}=0$. We have defined the trace operation so that it is basis-independent, and so, for simplicity, we can compute $\mathtt{\alpha}$, $\mathtt{\beta}$, and $\mathtt{\delta}$ in terms of the standard orthonormal basis $\{e_{1},\dots,e_{d}\}$ admitting $e_{r}-\overline{e_{r}}=0$ $\forall r\in\{1,\dots,d\}$. On that view, one immediately sees that $\mathtt{\alpha}=\mathtt{\alpha}'$. Next, we observe that $j\Lambda_{2}j=-\overline{\Lambda_{2}}$ and $j\Gamma_{2}j=-\overline{\Gamma_{2}}$. Therefore $\mathtt{\beta}=\mathtt{\beta}'$. Finally, we observe that $j\Lambda_{1}=\overline{\Lambda_{1}}j$ and $j\Gamma_{1}=\overline{\Gamma_{1}}j$, and after some algebra one finds that $\mathtt{\delta}=0$, finishing the proof. 

\subsection{\label{sec:level11}Simulating Generalized Measurements}
Equipped with \eqref{claim}, we are now ready to prove that generalized measurements in \textsc{qqt} can be simulated by usual quantum measurements in \textsc{cqm} with positive operator valued measures on complex Hilbert spaces. Let $\rho\in M_{d,d}(\mathbb{H})_{sa}$ be a quaternionic quantum state for a physical system $\mathfrak{S}$, and let $\mathfrak{M}_{\mathbb{H}}=\{E_{1},\dots ,E_{m}\}\subseteq\mathcal{E}(\mathbb{H}^{d})$ define a generalized quaternionic quantum measurement with outcome probabilities $p(r)=\mathrm{tr}(E_{r}\rho)$. Then, there exists a complex quantum state $\sigma(\rho)=\textstyle{\frac{1}{2}}\psi_{d,d}(\rho)\in M_{2d,2d}(\mathbb{C})_{sa}$ and a positive operator valued measure $\mathfrak{M}_{\mathbb{C}}=\{\psi_{d,d}(E_{1}),\dots ,\psi_{d,d}(E_{m})\}\subseteq\mathcal{E}(\mathbb{C}^{2d})$ on complex Hilbert space with outcome probabilities $q(r)=\mathrm{tr}\big(\psi_{d,d}(E_{r})\sigma(\rho)\big)$, such that $\forall r$: $q(r)=p(r)$.

For the proof, we begin by showing that $\psi_{d,d}$ preserves positive semi-definiteness. The spectral decomposition of positive semi-definite $\rho\in M_{d,d}(\mathbb{H})_{sa}$ is given by $\rho=\sum_{r=1}^{d}|\xi_{r}\rangle\lambda_{r}\langle\xi_{r}|$ in terms of $\lambda_{r}\in\mathbb{R}_{+}$ and eigenprojectors $\Xi_{r}=|\xi_{r}\rangle\langle\xi_{r}|$. We have that $\psi_{d,d}$ is $\mathbb{R}-$linear from \eqref{rLin}, and from \eqref{homo} it is clear that $\psi_{d,d}$  maps projections on $\mathbb{H}^{d}$ to projections on $\mathbb{C}^{2d}$. Thus, $\psi_{d,d}(\rho)$ is a positive semi-definite operator on complex Hilbert space. Next, we define positive semi-definite $\sigma(\rho)=\textstyle{\frac{1}{2}}\psi_{d,d}(\rho)$, and by \eqref{claim} we have that $\sigma(\rho)$ is unit-trace. Therefore, $\sigma(\rho)$ is a valid complex quantum state. Also, from the definition of $\psi_{d,d}$ it follows that $\psi_{d,d}(\mathds{1}_{\mathbb{H}^{d}})=\mathds{1}_{\mathbb{C}^{2d}}$, and applying $\mathbb{R}-$linearity of $\psi_{d,d}$ once again, it follows that $\mathfrak{M}_{\mathbb{C}}=\{\psi_{d,d}(E_{1}),\dots,\psi_{d,d}(E_{m})\}$ is a valid positive operator valued measure on complex Hilbert space. Finally, applying the quaternionic Born rule \eqref{bornRule} and using \eqref{claim} we see that $\forall r$:
\begin{equation}
p(r)=\mathrm{tr}(E_{r}\rho)=\mathrm{tr}\big(\psi_{d,d}(E_{r})\sigma(\rho)\big)=q(r)\text{,}
\end{equation} 
finishing the proof.

\subsection{\label{sec:level12}Simulating Quantum Channels}
In this section, we prove that quaternionic quantum channels can be simulated by completely positive trace preserving maps in usual \textsc{cqm}. Let $\rho\in M_{d,d}(\mathbb{H})_{sa}$ be a quaternionic quantum state for a physical system $\mathfrak{S}$, and let $\Phi:M_{d,d}(\mathbb{H})_{sa}\rightarrow M_{p,p}(\mathbb{H})_{sa}$ be a quaternionic quantum channel whose action is defined via 
\begin{equation}
\Phi(\rho)=\sum_{r=1}^{n}A_{r}\rho A_{r}^{*}\text{,}
\end{equation}
where $A_{r}\in M_{p,d}(\mathbb{H})$ and $\sum_{r=1}^{n}A_{r}A_{r}^{*}=\mathds{1}_{\mathbb{H}^{p}}$. Then, there exists a complex quantum channel $\Theta$ whose action on $\sigma(\rho)$ is defined via 
\begin{equation}
\Theta(\sigma(\rho))=\sum_{r=1}^{n}\psi_{p,d}(A_{r})\sigma(\rho)\psi_{p,d}(A_{r})^{*}\text{,}
\end{equation} 
and given an arbitrary quaternionic quantum measurement defined by $\mathfrak{M}_{\mathbb{H}}=\{E_{1},\dots ,E_{m}\}\subseteq\mathcal{E}(\mathbb{H}^{d})$ one has that $\forall r$:
\begin{equation}
\mathrm{tr}\Big(E_{r}\Phi\big(\rho\big)\Big)=\mathrm{tr}\Big(\psi_{p,p}\big(E_{r}\big)\Theta\big(\sigma(\rho)\big)\Big)\text{,}
\label{equiv}
\end{equation}
\\
where $\mathfrak{M}_{\mathbb{C}}=\{\psi_{p,p}(E_{1}),\dots,\psi_{p,p}(E_{m})\}$ is a positive operator valued measure on complex Hilbert space. Put otherwise, any generalized preparation $\rightarrow$ transformation $\rightarrow$ measurement process in \textsc{qqt} corresponds to an algorithm in usual complex quantum information theory.

For the proof, note that \eqref{rLin}, \eqref{homo}, and \eqref{star} imply that $\sum_{r}\psi_{p,d}(A_{r})\psi_{p,d}(A_{r})^{*}=\mathds{1}_{\mathbb{C}^{2p}}$, so $\Theta$ is a valid complex quantum channel. We have already established that $\sigma(\rho)$ is a valid complex quantum state, and so it follows that $\Theta(\sigma(\rho))$ is a valid complex quantum state. Now, again using \eqref{rLin}, \eqref{homo}, and \eqref{star} we have that
\begin{equation}
\Theta(\sigma(\rho))=\textstyle{\frac{1}{2}}\psi_{p,p}\Big(\sum_{r}A_{r}\rho A_{r}^{*}\Big)\text{,}
\end{equation} 
and so by \eqref{claim} we see that \eqref{equiv} holds, finishing the proof.

\section{\label{sec:level13}Conclusion}
Ultimately, one would like to use the developing technologies of quantum information science to test the validity of usual \textsc{cqm} versus \textsc{qqt} \cite*{Peres79}. Before one can perform such tests, however, one must have a clear conception of the relations and contrasts between these two theories with respect to the full apparatus of quantum information theory, not just the projective measurements and unitary operations considered in \cite*{Fernandez03}. This paper fills that gap in the literature. We have shown that \textit{all} generalized quantum dynamics in \textsc{qqt} can be realized as usual quantum dynamics in \textsc{cqm}. In particular, we have shown that generalized measurements associated with positive operator valued measures on quaternionic modules can be simulated by usual quantum measurements in \textsc{cqm}. Furthermore, we have shown that quaternionic quantum channels can be simulated by completely positive trace preserving maps in \textsc{cqm}. These results offer a new vantage point to view quaternionic quantum algorithms from \textit{inside} usual complex quantum information theory.

\section{\label{sec:level14}Acknowledgments}
The author thanks Chris Fuchs for his guidance and support. The author also thanks Howard Barnum for discussions on Jordan algebras, and \AA sa Ericsson for feedback on the manuscript. This work was supported in part by the U. S. Office of Naval Research (Grant No. N00014-09-1-0247), and by the province of Ontario through OGS.

\bibliography{qqdchs}

\end{document}